\documentclass{article}

\usepackage[preprint]{neurips_2026}
\usepackage[utf8]{inputenc}
\usepackage[T1]{fontenc}
\usepackage{hyperref}
\usepackage{url}
\usepackage{booktabs}
\usepackage{graphicx}
\usepackage{float}
\usepackage{placeins}
\usepackage{amsfonts}
\usepackage{amsmath}
\usepackage{amsthm}
\usepackage{amssymb}
\usepackage{enumitem}
\usepackage{nicefrac}
\usepackage{microtype}
\usepackage{xcolor}
\usepackage{multirow}
\usepackage{array}
\usepackage{tabularx}
\usepackage{algorithm}
\usepackage{algpseudocode}
\usepackage[most]{tcolorbox}

\newcolumntype{Y}{>{\raggedright\arraybackslash}X}

\definecolor{agentcardframe}{HTML}{B56464}
\definecolor{agentcardtitle}{HTML}{F1CCC8}
\definecolor{skillcardframe}{HTML}{6D9A73}
\definecolor{skillcardtitle}{HTML}{DCEFD9}
\definecolor{provcardframe}{HTML}{B58A4B}
\definecolor{provcardtitle}{HTML}{F3DEB7}

\newtcolorbox{agentcard}[1]{
  enhanced,
  breakable,
  colback=white,
  colframe=agentcardframe,
  colbacktitle=agentcardtitle,
  coltitle=black,
  fonttitle=\bfseries,
  title={#1},
  boxrule=0.7pt,
  arc=2pt,
  left=6pt,
  right=6pt,
  top=5pt,
  bottom=5pt
}

\newtcolorbox{skillcard}[1]{
  enhanced,
  breakable,
  colback=white,
  colframe=skillcardframe,
  colbacktitle=skillcardtitle,
  coltitle=black,
  fonttitle=\bfseries,
  title={#1},
  boxrule=0.7pt,
  arc=2pt,
  left=6pt,
  right=6pt,
  top=5pt,
  bottom=5pt
}

\newtcolorbox{provcard}[1]{
  enhanced,
  breakable,
  colback=white,
  colframe=provcardframe,
  colbacktitle=provcardtitle,
  coltitle=black,
  fonttitle=\bfseries,
  title={#1},
  boxrule=0.7pt,
  arc=2pt,
  left=6pt,
  right=6pt,
  top=5pt,
  bottom=5pt
}

\title{SkillMAS: Skill Co-Evolution with LLM-based Multi-Agent System}

\author{%
\bfseries
Shuai Pan\textsuperscript{1,*},
Yixiang Liu\textsuperscript{1,*},
Jiaye Gao\textsuperscript{1},
Te Gao\textsuperscript{2},
Weiwen Liu\textsuperscript{1,\textdagger},\\
\bfseries
Jianghao Lin\textsuperscript{1,\textdagger},
Zhihui Fu\textsuperscript{3},
Jun Wang\textsuperscript{3,\textdagger},
Weinan Zhang\textsuperscript{1,\textdagger},
Yong Yu\textsuperscript{1}\\
\normalfont
\textsuperscript{1}Shanghai Jiao Tong University \quad
\textsuperscript{2}Central South University \quad
\textsuperscript{3}OPPO\\
\normalfont
\textsuperscript{*}Equal contribution. \quad
\textsuperscript{\textdagger}Corresponding authors.
}

\begin{document}

\maketitle
\makeatletter
\if@anonymous\else
\begingroup
\renewcommand{\thefootnote}{\textdagger}
\footnotetext{Correspondence: \texttt{\{wwliu, linjianghao, wnzhang\}@sjtu.edu.cn}, \texttt{wangjun15@oppo.com}.}
\addtocounter{footnote}{-1}
\endgroup
\fi
\makeatother

\begin{abstract}
Large language model (LLM) agent systems are increasingly expected to improve after deployment, but existing work often decouples two adaptation targets: skill evolution and multi-agent system (MAS) restructuring. This separation can create organization bottlenecks, context pressure, and mis-specialization. We present \textbf{SkillMAS}, a non-parametric framework for adaptive specialization in multi-agent systems that couples skill evolution with MAS restructuring. SkillMAS uses Utility Learning to assign credit from verified execution traces, bounded skill evolution to refine reusable procedures without unfiltered library growth, and evidence-gated MAS restructuring when retained failures and Executor Utility indicate a structural mismatch. Across embodied manipulation, command-line execution, and retail workflows, SkillMAS is competitive under the reported harnesses while clarifying how post-deployment specialization is attributed, updated, and applied.
\end{abstract}

\section{Introduction}

LLM agents are increasingly expected not only to complete long-horizon tasks, but also to improve after deployment as they accumulate verified execution traces, reusable procedures, and coordination experience. This creates a system-level problem: a deployed agent stack must decide what to retain, what to revise, and when a fixed collaboration structure has itself become the bottleneck. Prior work on multi-agent orchestration, externalized agent infrastructure, post-deployment memories or skills, and Agentic ROI suggests that practical agent systems should keep improving after deployment rather than remain fixed pipelines \citep{hong2023metagpt,wu2023autogen,zhou2026externalization,ouyang2025reasoningbank,xia2026skillrl,wang2026autoagent,liu2025agenticroi}.

Existing work still treats two tightly coupled adaptation targets as largely separate problems: skill evolution and MAS restructuring. One line adapts MAS organization through spawning, orchestration, role/profile updates, interaction rewards, or topology selection \citep{costa2026agentspawn,yu2026adaptorch,ma2025agenticnn,xue2025comas,lu2024morphagent,nie2026skillgraph}. A separate line evolves reusable skills or distills trajectory-local lessons into transferable procedures, typically without simultaneously editing executor boundaries \citep{zhang2026memskill,xia2026skillrl,ni2026trace2skill,zhang2026coevoskills,alzubi2026evoskill}. In practice, however, these targets interact directly: skill evolution changes what MAS organization must route and maintain, while MAS organization determines whether evolved skills can be reused without excessive context load or responsibility ambiguity. We call this system-level mismatch \textbf{adaptation decoupling}: skill evolution and MAS restructuring are optimized separately even though each changes the operating conditions of the other.

\begin{figure}[t]
\centering
\includegraphics[width=\textwidth]{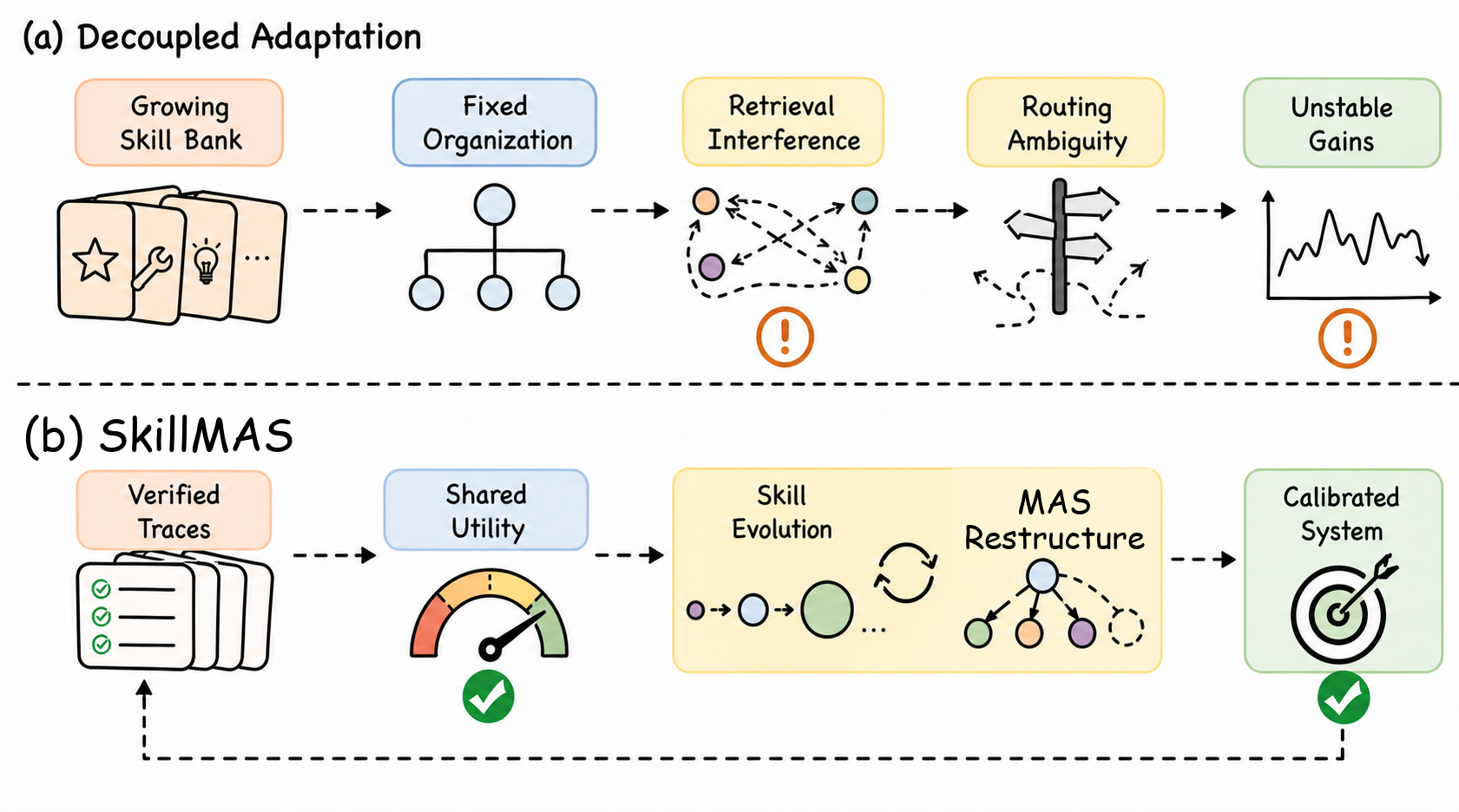}
\caption{Coupled adaptation is the paper's central motivation: decoupled skill evolution can increase interference under fixed MAS organization, whereas SkillMAS coordinates skill evolution and MAS restructuring from shared verified traces.}
\label{fig:motivation}
\end{figure}

Adaptation decoupling creates three recurring failure modes. Verified execution trace reuse can introduce redundant or low-value evidence for skill evolution; skill evolution can make MAS organization harder to route and maintain; and fixed MAS organization can lag behind changing task structure or runtime complexity. We therefore make three design requirements explicit: credit should be assigned from verified execution traces rather than retrieval alone; skill evolution should avoid overwhelming MAS organization; and skill evolution should be comparable with MAS restructuring under the same empirical evidence. Without a shared evidence surface, the system cannot tell whether the next useful intervention is skill evolution, MAS restructuring, or no change at all. Figure~\ref{fig:motivation} illustrates this failure mode.

To this end, we introduce \textbf{SkillMAS}, a non-parametric framework that treats skill evolution and MAS restructuring as one empirical loop. Utility Learning updates Skill Utility and Executor Utility only from verified execution traces, bounded skill evolution changes the skill library without accepting every local patch, and MAS restructuring changes MAS organization only when retained failures and Executor Utility indicate an organizational bottleneck. The contribution is not that either update target is new in isolation, but that both are constrained by one verified-trace evidence surface. The empirical section evaluates this scaffold through benchmark performance, an ALFWorld stress test, and round-by-round adaptation trajectories.

We make three contributions:
\begin{itemize}[leftmargin=*,topsep=2pt,itemsep=0pt,parsep=0pt]
\item We formulate adaptation decoupling as a system-level problem in post-deployment specialization of self-evolving MAS.
\item We introduce SkillMAS, a non-parametric scaffold that places Utility Learning, bounded skill evolution, and evidence-gated MAS restructuring under shared verified-trace evidence.
\item We evaluate this scaffold across embodied manipulation, command-line OS workflows, and retail-service interaction.
\end{itemize}

\section{Method}
\label{sec:method}

\begin{figure}[t]
\centering
\includegraphics[width=\textwidth]{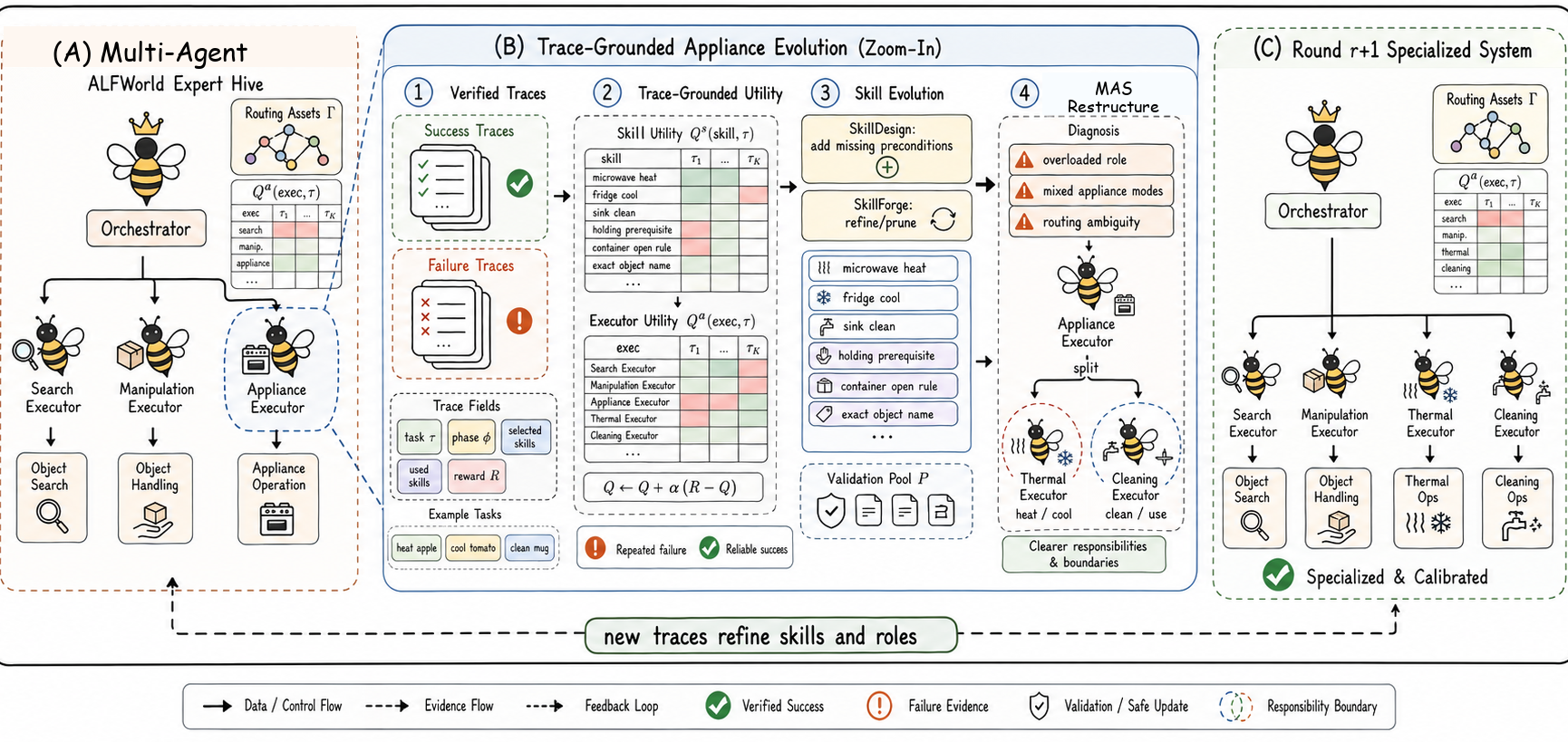}
\caption{SkillMAS uses one retained evidence set to update utilities, repair skills, and justify MAS restructuring only when Executor Utility exposes overloaded MAS organization.}
\label{fig:method_pipeline}
\end{figure}

SkillMAS addresses adaptation decoupling without retraining the underlying language models. One adaptation round executes a batch of episodes, learns utility from verified execution traces, constructs a retained evidence set, and then applies skill evolution together with evidence-gated MAS restructuring. Figure~\ref{fig:method_pipeline} gives an ALFWorld example of how the same retained evidence coordinates skill evolution and MAS restructuring.

\subsection{Round State and Adaptation Loop}
\label{sec:multi-cycle-mas}

SkillMAS maintains the following round state:
\begin{equation}
X_r=\bigl(\mathcal{L}_r,\mathcal{A}_r,Q^s_r,Q^a_r,\mathcal{P}_r,\mathcal{K}_r\bigr),
\label{eq:round_state}
\end{equation}
where $\mathcal{L}_r$ is the skill library, $\mathcal{A}_r$ the executor set, $Q^s_r$ and $Q^a_r$ the Skill Utility and Executor Utility tables, $\mathcal{P}_r$ a validation pool for new or heavily revised skills, and $\mathcal{K}_r$ a policy-derived expert policy index.

With $X_r$ fixed, one adaptation round begins by executing the current system:
\begin{equation}
\mathcal{T}_r
= \mathrm{Exec}(X_r).
\label{eq:exec_round}
\end{equation}
Here $n_r$ is the number of episodes executed in round $r$, and we write $\xi\in\mathcal{T}_r$ for an episode-level verified trace from that round. Each $\xi$ records the attempted task, selected skills, routed executors, produced trajectory, and verified terminal outcome. When we need the executor-local portion associated with a routed executor $a$, we write $\xi(a)$ for the corresponding executor trace slice inside the same episode trace.

The trace set is then converted into post-execution utility estimates:
\begin{equation}
\bigl(Q^{s,+}_r,Q^{a,+}_r\bigr)
= \mathrm{Learn}\!\left(Q^s_r,Q^a_r,\mathcal{T}_r\right).
\label{eq:learn_round}
\end{equation}
The superscript $+$ denotes utilities after Utility Learning but before the coupled adaptation update.

SkillMAS does not pass the entire trace set to the update modules. It first constructs a retained evidence set:
\begin{equation}
\widetilde{\mathcal{T}}_r
= \mathrm{Retain}\!\left(\mathcal{T}_r,Q^{s,+}_r,Q^{a,+}_r\right)
\subseteq \mathcal{T}_r.
\label{eq:retained_evidence}
\end{equation}
This lightweight filtering step is not a learned controller and is not the main contribution. It retains verified traces that are most useful for adaptation, such as repeated failures, near misses, reusable successes, and retrieval/execution mismatches. The retained evidence set $\widetilde{\mathcal{T}}_r$ then drives one coupled adaptation update: skill evolution proposes bounded changes to $\mathcal{L}_r$ and $\mathcal{P}_r$, while MAS restructuring keeps, adds, merges/removes, or modifies executors when the evidence supports a structural mismatch. The accepted skill evolution and MAS restructuring changes jointly produce $X_{r+1}$.
Appendix Algorithm~\ref{alg:skillmas_round} summarizes the concrete round-level contract used in the experiments, and Appendix Table~\ref{tab:heuristic_instantiations} lists the benchmark-local heuristics that instantiate retained-evidence construction, diagnosability, and restructuring predicates.

\subsection{Utility Learning}
\label{subsec:trace_grounded_utility}

SkillMAS learns Skill Utility and Executor Utility from task execution outcomes grounded in verified execution traces, rather than from retrieval alone. Equation~\ref{eq:learn_round} is instantiated over $\mathcal{T}_r$, where each element $\xi$ is an episode-level verified trace and each $\xi(a)$ denotes the executor-local slice created during a delegated step. Utility updates are grounded in what was actually executed and validated.

Let $\tau$ denote the episode-level task type. SkillMAS maintains two utility estimates for two decision layers: skill selection and executor selection:
\begin{align}
Q^s_r(s,\tau) &: \text{Skill Utility for skill } s \text{ on task type } \tau,\\
Q^a_r(a,\tau) &: \text{Executor Utility for executor } a \text{ under task type } \tau.
\end{align}
$Q^s_r$ is used for task-conditioned skill selection from the skill library, while $Q^a_r$ ranks candidate executors under the same task type. Since ALFWorld success is binary, we use $R(\xi)\in\{0,1\}$ and maintain $Q^s_r,Q^a_r\in[0,1]$ as empirical success-probability estimates under their conditions. SkillMAS decouples candidate retrieval from utility attribution: retrieval proposes potentially relevant skills, but only execution-supported skills receive positive credit.

For an episode-level trace $\xi\in\mathcal{T}_r$ and executor $a$, let $\mathcal{S}^{\mathrm{sel}}_\xi(a)$ be the selected skill set. SkillMAS defines the used skill set as the selected skills that are actually invoked by the executor or supported by the executor-local trace slice:
\begin{equation}
\mathcal{S}^{\mathrm{used}}_\xi(a)
=
\left\{
 s \in \mathcal{S}^{\mathrm{sel}}_\xi(a)
\;\middle|\;
 s \text{ is invoked or execution-supported in } \xi(a)
\right\}.
\label{eq:trace_grounded_credit}
\end{equation}

Concretely, each executor-local trace slice $\xi(a)$ stores the delegated subtask, selected skill identifiers, action log or tool calls, and verifier-backed terminal outcome. A selected skill is counted as \emph{used} only when the executor explicitly invokes its identifier or the realized trace matches its guard-and-step pattern under the benchmark verifier. This is a conservative operational attribution rule rather than causal identification; its purpose is to avoid assigning positive credit to merely retrieved skills.

For each trace $\xi\in\mathcal{T}_r$, SkillMAS applies one Monte Carlo credit-assignment rule over two update domains:
\begin{equation}
\mathcal{U}^{s}_\xi(a)=\mathcal{S}^{\mathrm{used}}_\xi(a),
\qquad
\mathcal{U}^{a}_\xi=\mathcal{A}^{\mathrm{exec}}_\xi,
\label{eq:trace_update_domain}
\end{equation}
\begin{equation}
Q^{\ell,+}_r(x,z)
\leftarrow Q^\ell_r(x,z)+\alpha_{r,x,z}\bigl(R(\xi)-Q^\ell_r(x,z)\bigr),
\qquad \ell\in\{s,a\}.
\label{eq:trace_mc_update}
\end{equation}
For Skill Utility, $x=s$ and $z=\tau$ with $s\in\mathcal{U}^{s}_\xi(a)$. For Executor Utility, $x=a$ and $z=\tau$ with $a\in\mathcal{U}^{a}_\xi$. Here $\alpha_{r,x,z}\in(0,1]$ is the update rate, and $\mathcal{A}^{\mathrm{exec}}_\xi$ is the set of executors with verified execution traces in $\xi$. In the current benchmark instantiation, we use the count-based schedule
\begin{equation}
\alpha_{r,x,z}=\frac{1}{1+N_r(x,z)},
\label{eq:count_step_size}
\end{equation}
where $N_r(x,z)$ is the number of prior verified updates for the same utility entry. Unseen entries therefore receive $\alpha=1$, and repeated evidence decays the update rate automatically without introducing a second optimizer or benchmark-specific momentum term. This keeps the learning rule compact while preserving the paper's main point: skill credit is restricted to execution-supported skills, whereas executor credit is assigned to the routed executors that actually participated. Overall, similarity retrieves candidates, verified execution traces determine usage, and outcomes define the learning target.

These utility tables are empirical summaries of verified outcomes, not convergence guarantees. They are used to rank candidate skills and executors under the current trace distribution, and their values can move when the task mix, skill library, or MAS organization changes.

\subsection{Skill Evolution}
\label{sec:skill-evolution}

Skill evolution converts retained evidence into bounded edits to the skill library. SkillMAS treats a skill as an agent-native package containing applicability conditions, procedural steps, failure guards, and verification checks. Clean success traces can yield reusable motifs, while failed traces enter the patch pool only when bounded analysis identifies one dominant editable cause, such as a missing precondition, incorrect action order, misleading retrieval match, skill conflict, or bad executor assignment. Formally, bounded diagnosability maps a retained failure trace to
\begin{equation}
D_r(\xi)=\bigl(c(\xi),u(\xi),b(\xi)\bigr)\in \mathcal{C}\times\{0,1\}\times\mathcal{B},
\label{eq:diagnosability}
\end{equation}
where $c(\xi)$ is the dominant cause, $u(\xi)=1$ indicates that the cause is uniquely identifiable from the trace, and $b(\xi)\in\mathcal{B}$ is the bounded update tag used to route the trace in the current benchmark instantiation:
\[
\mathcal{B}=
\left\{
\begin{array}{l}
\texttt{add-guard},\ \texttt{reorder-step},\ \texttt{tighten-retrieval},\\
\texttt{split-skill},\ \texttt{handoff-to-structure},\ \varnothing
\end{array}
\right\}.
\]
The first four tags denote bounded local repair categories. The tag \texttt{handoff-to-structure} routes the trace to structural analysis rather than direct skill repair, and $\varnothing$ means that no bounded edit is proposed. A failure is treated as locally diagnosable for skill repair only when $u(\xi)=1$ and $b(\xi)\in\{\texttt{add-guard},\texttt{reorder-step},\texttt{tighten-retrieval},\texttt{split-skill}\}$. Before synthesis or repair, SkillMAS retrieves a small set of policy cards from $\mathcal{K}_r$, a fixed index over seed skills, previously validated skills, and benchmark-local expert exemplars. $\mathcal{K}_r$ supplies repair priors rather than a general planner. Let $y(\xi)\in\{0,1\}$ denote the verified outcome. Each verified execution trace yields at most one local proposal:
\begin{equation}
p(\xi)=
\begin{cases}
A^{+}(\xi), & y(\xi)=1,\\
A^{-}(\xi), & y(\xi)=0 \text{ and diagnosable},\\
\varnothing, & \text{otherwise}.
\end{cases}
\label{eq:patch_proposal}
\end{equation}
Here $A^{+}(\xi)$ denotes a success-derived skill proposal, such as a reusable motif extracted from a verified successful trace, and $A^{-}(\xi)$ denotes a failure-derived local repair proposal consistent with $b(\xi)$, such as adding a guard, reordering steps, tightening retrieval scope, or splitting an overloaded skill.

Local proposals are then consolidated:
\begin{equation}
\Delta^{\mathrm{skill}}_r=
\mathrm{SkillEvolve}\!\left(\{p(\xi)\}_{\xi\in\widetilde{\mathcal{T}}_r},\mathcal{L}_r,\mathcal{K}_r\right).
\label{eq:skill_merge_update}
\end{equation}
This stage is trace-driven: it deduplicates proposals against the existing library and policy cards, refines weak skills into narrower patches, prunes low-value or redundant skills, and applies penalties after round-level performance drops. Its bounded action set is $\{\texttt{create},\texttt{refine},\texttt{prune},\texttt{hold-in-pool},\texttt{no-op}\}$, applied at most once per implicated skill cluster in a round. Newly created or heavily rewritten skills remain in $\mathcal{P}_r$ until later rounds provide sufficient usage evidence.

\subsection{Evidence-Gated MAS Restructuring}
\label{sec:agent-scaling}

Evidence-Gated MAS Restructuring addresses failures that are not resolved by skill evolution alone. As $|\mathcal{L}_r|$ grows, fixed MAS organization faces more retrieval choices, more selection interference, and greater coordination burden. SkillMAS therefore changes $\mathcal{A}_r$ only when retained evidence and Executor Utility show that the remaining problem is overloaded or poorly separated MAS organization. The system first refines retained evidence into structured diagnostic artifacts:
\begin{equation}
\mathcal{F}_r = H_r\!\left(\widetilde{\mathcal{T}}_r,\mathcal{L}_r,\mathcal{A}_r,Q^{a,+}_r, \Delta^{\mathrm{skill}}_r\right).
\label{eq:failure_artifacts}
\end{equation}
These artifacts summarize the failure pattern, implicated executors, Executor Utility evidence, skill overlap, and pending skill update. Typical structural evidence includes high uncertainty among similar executors, repeated failures concentrated in a task family after skill repair, or a broad executor owning mutually interfering skill clusters. Based on $\mathcal{F}_r$, the system proposes a bounded MAS restructuring decision:
\begin{equation}
\Delta^{\mathrm{agent}}_r = G_r\!\left(\mathcal{F}_r, \mathcal{A}_r, Q^{a,+}_r\right).
\label{eq:agent_update}
\end{equation}
The operator $G_r$ returns one of four outcomes: keep the current MAS organization, add a specialist, remove or merge a redundant executor, or modify responsibility boundaries and skill ownership. It returns no edit unless the retained evidence supports a structural mismatch, and at most one restructuring edit is applied in a round. The bounded action set of $G_r$ is $\{\texttt{keep},\texttt{add},\texttt{merge/remove},\texttt{modify}\}$; its predicates are fixed benchmark-local rules over cluster mass, utility gap, and owned-skill overlap, not universal thresholds. When MAS restructuring is triggered, SkillMAS transfers validated skills, consolidates redundant skill ownership, or narrows executor prompts so that each executor covers a clearer task region than before the update.

\section{Experiments}
\label{sec:experiments}

We evaluate SkillMAS on embodied household manipulation, command-line OS workflows, and retail-service interaction. The experiment section separates contextual benchmark performance from mechanism evidence and adaptation-process interpretation.

\subsection{Benchmark and Evaluation Setup}

All comparisons use the harness and model assignment available for each benchmark domain. We use GPT-4o-mini wherever possible. On the ALFWorld unseen domain \citep{shridhar2021alfworld}, adaptation is run on a fixed 70-task train subset selected once by task-family-proportional sampling from the 134-task evolution pool. The full 134-task unseen split is reserved for evaluating selected checkpoints, and the best round on the 70-task train subset in Table~\ref{tab:process_rounds} is the checkpoint that reaches 126/134 = 94.0\% on the full unseen evaluation. On the Lifelong Agent Bench OS Task domain \citep{zheng2025lifelongagentbench}, we use a fixed 100-task train subset for round-by-round process tracking and report the full 150-task evaluation score from the best checkpoint. On the $\tau$-Bench Retail domain \citep{yao2024taubench}, adaptation and development use the official 74-task train subset, while the main benchmark evaluation is the base set with 114 tasks total: 74 train tasks plus 40 test tasks. The benchmark fixes two model roles, a GPT-4.1-mini executor and a GPT-4.1-2025-04-14 user simulator. Appendix Table~\ref{tab:experiment_config} summarizes these adaptation/evaluation splits and model-role assignments. We report success rate as the primary metric and use active skill count, active executor count, and task-family success to interpret SkillMAS's adaptation process.

Table~\ref{tab:main_results} is a contextual comparison rather than a protocol-matched leaderboard. The ALFWorld Direct LLM and ReAct entries are published reference numbers from ReflAct Table~2 \citep{kim2025reflact}; CDMem and Traj-Bootstrap are taken from their original papers \citep{gao2025cdmem,sarukkai2025trajbootstrap}. We mark cross-paper ALFWorld entries with ``Ref.'' because protocol drift could matter, especially for small margins.

\subsection{Overall Performance}

Table~\ref{tab:main_results} summarizes success rates across the reported benchmark settings.

\begin{table}[H]
\centering
\small
\setlength{\tabcolsep}{9pt}
\caption{Success rates (\%). Ref. entries are contextual published scores; unmarked entries are our evaluations or reruns, not a protocol-matched leaderboard.}
\label{tab:main_results}
\begin{tabular}{@{}lccc@{}}
\toprule
\textbf{Method} & \textbf{ALFWorld} & \textbf{Lifelong Agent Bench} & \textbf{$\tau$-Bench} \\
 & \textbf{Unseen domain} & \textbf{OS Task domain} & \textbf{Retail domain} \\
\midrule
Direct LLM & 43.3 (Ref.) & 59.3 & 61.4 \\
ReAct \citep{yao2023react} & 53.0 (Ref.) & 62.0 & 62.3 \\
CDMem \citep{gao2025cdmem} & 85.8 (Ref.) & 68.0 & 68.4 \\
Traj-Bootstrap \citep{sarukkai2025trajbootstrap} & 93.0 (Ref.) & 70.0 & 68.4 \\
\midrule
\textbf{SkillMAS} & \textbf{94.0} & \textbf{76.7} & \textbf{70.2} \\
\bottomrule
\end{tabular}
\end{table}

Across the three reported benchmarks, SkillMAS attains the highest success rate among the methods shown in Table~\ref{tab:main_results}. On ALFWorld, SkillMAS reaches 94.0\% on the full unseen split, competitive with the strongest published reference score of 93.0\% under this contextual comparison. On Lifelong Agent Bench OS Task, SkillMAS reaches 76.7\%, compared with 70.0\% for our Traj-Bootstrap rerun and 68.0\% for CDMem. On $\tau$-Bench Retail, the selected checkpoint reaches 70.2\% on the official base set.

\subsection{Ablation Experiment}

The current archive does not include controlled frozen-target retraining ablations for ALFWorld. Table~\ref{tab:stress_test} therefore reports an ALFWorld-only mechanism stress test that transplants the final MAS organization or final skill library into a partially frozen counterpart. It probes mismatch and overload, but it is not a protocol-matched causal factorization.

\begin{table}[H]
\centering
\caption{ALFWorld transplant stress test for skill evolution and MAS organization mismatch.}
\label{tab:stress_test}
\small
\setlength{\tabcolsep}{4pt}
\begin{tabular}{@{}lccc@{}}
\toprule
\textbf{Variant} & \textbf{Skill evolution} & \textbf{MAS restructuring} & \textbf{Success} \\
\midrule
Full SkillMAS & \checkmark & \checkmark & 126/134 (94.0\%) \\
Final library, seed MAS organization & \checkmark & $\times$ & 92/134 (68.7\%) \\
Specialized MAS organization, seed skills & $\times$ & \checkmark & 67/134 (50.0\%) \\
Seed baseline & $\times$ & $\times$ & 102/134 (76.1\%) \\
\bottomrule
\end{tabular}
\vspace{-0.7em}
\end{table}

\subsection{ALFWorld Task-Family Breakdown}

The ALFWorld task-family breakdown in Table~\ref{tab:alfworld_task_type} shows where adaptation helps. The largest absolute gain is in \texttt{examine}, rising from 5/18 to 17/18, which matches the introduction of search- and examination-oriented specialization. Other task families start from stronger seed baselines and show smaller but consistent gains. We treat task-type success as the externally interpretable evidence and raw Executor Utility values as internal adaptation state.

\begin{table}[H]
\centering
\caption{ALFWorld task-type success on the 134-task unseen split.}
\label{tab:alfworld_task_type}
\begin{tabular}{@{}lccc@{}}
\toprule
\textbf{Task type} & \textbf{Seed round} & \textbf{Best round} & \textbf{Gain} \\
\midrule
\texttt{examine} & 5/18 (27.8\%) & 17/18 (94.4\%) & +12 \\
\texttt{pick\_and\_place} & 28/34 (82.4\%) & 33/34 (97.1\%) & +5 \\
\texttt{pick\_clean\_then\_place} & 27/31 (87.1\%) & 28/31 (90.3\%) & +1 \\
\texttt{pick\_cool\_then\_place} & 18/21 (85.7\%) & 20/21 (95.2\%) & +2 \\
\texttt{pick\_heat\_then\_place} & 11/13 (84.6\%) & 12/13 (92.3\%) & +1 \\
\texttt{pick\_two\_obj\_and\_place} & 13/17 (76.5\%) & 16/17 (94.1\%) & +3 \\
\bottomrule
\end{tabular}
\end{table}

\section{Analysis}
\label{sec:analysis}

\subsection{Mechanism Evidence and Scope}

Table~\ref{tab:stress_test} is consistent with the narrower point that mismatch between skill evolution and MAS organization can hurt performance. Injecting the final skill library into seed MAS organization falls to 92/134, below the 102/134 seed baseline, suggesting selection and application overload. Keeping specialized MAS organization but reverting to seed skills falls further to 67/134, suggesting that MAS restructuring without matching skill evolution is also brittle. Since the archive lacks controlled frozen-target retraining ablations, we treat these results as an ALFWorld-specific stress test rather than a factorized estimate of causal contribution. The asymmetry between 92/134 and 67/134 is itself informative: one-sided skill growth can overload a fixed routing structure, while one-sided structural specialization can leave the system without enough validated procedural content to exploit that structure.

\subsection{Adaptation Trajectories}

Appendix Table~\ref{tab:process_rounds} records skill evolution and MAS restructure process. ALFWorld grows from 5 to 13 active skills while train success rises from 54/70 to 66/70, and the selected round-5 checkpoint transfers to 126/134 on the full unseen split. Lifelong Agent Bench OS Task shows the complementary point: active skills reach 10 by round 3, where success peaks at 78/100, but round 4 keeps the same skill count and drops to 75/100. Skill count is therefore an interpretability signal, not the claim itself. The $\tau$-Bench trajectory shows the opposite structural pattern: the selected checkpoint keeps one executor while improving from 43/74 to 51/74, and expanding that checkpoint with an additional preflight executor drops train performance to 32/74. In this benchmark, the gain comes from better skill use under stable single-agent execution rather than from beneficial MAS expansion.

\subsection{ALFWorld Task-Family Insights}

Table~\ref{tab:alfworld_task_type} shows that the aggregate ALFWorld gain is not spread evenly across task families. The dominant improvement comes from \texttt{examine}, which rises from 5/18 to 17/18; this is the clearest evidence that the added search- and examination-oriented specialization fixes a concrete weakness in the seed system rather than merely smoothing already-strong cases. By contrast, \texttt{pick\_clean\_then\_place}, \texttt{pick\_cool\_then\_place}, and \texttt{pick\_heat\_then\_place} start from strong seed baselines and improve only modestly, suggesting that these families benefit more from incremental procedure sharpening than from major structural change. The intermediate gains on \texttt{pick\_and\_place} and \texttt{pick\_two\_obj\_and\_place} fit a third pattern: once search is more reliable, downstream manipulation families improve as a secondary effect because object and receptacle grounding become cleaner before action execution.

\section{Related Work}
\label{sec:related}

\paragraph{Multi-agent systems.}
LLM-based multi-agent systems use role-based collaboration, message passing, and orchestration to solve tasks that exceed a single agent's capacity \citep{hong2023metagpt,wu2023autogen}. Recent work treats MAS organization itself as adaptive: specialization can redistribute cognitive burden, homogeneous scaling can saturate, and systems can spawn agents, select topologies, update roles, or evolve decentralized agent profiles over time \citep{shang2025unitedminds,yang2026diversity,costa2026agentspawn,yu2026adaptorch,ma2025agenticnn,xue2025comas,lu2024morphagent}. SkillGraph is especially close because it co-evolves multimodal skills and collaboration topology, but it learns graph topology with a multimodal graph transformer, whereas SkillMAS uses non-parametric verified-trace utilities and bounded MAS restructuring \citep{nie2026skillgraph}. SkillMAS shifts the trigger from task complexity or interaction rewards alone to pressure created by skill evolution and observed in retained verified traces.

\paragraph{Self-evolving agents.}
A second line of work studies how agents improve after deployment by learning from experience. Reflexion and Voyager showed that reflective feedback and accumulated executable skills can improve future behavior without retraining from scratch \citep{shinn2023reflexion,wang2023voyager}. Recent systems make online adaptation more explicit through reasoning memories, exploration-aware self-improvement, textual backpropagation, memory meta-evolution, joint skill-policy evolution, and elastic memory orchestration \citep{ouyang2025reasoningbank,fang-etal-2025-webevolver,ma2025agenticnn,zhang2025memevolve,xia2026metaclaw,wang2026autoagent}. These methods establish that deployed agents should not remain fixed; SkillMAS focuses on when evolving reusable knowledge should also change executor boundaries.

\paragraph{Skill and memory learning.}
The closest skill-layer work studies utility-aware memory usage and trajectory-grounded skill learning. MemRL formalizes memory reuse as non-parametric reinforcement learning over episodic memory; MemSkill, SkillRL, Trace2Skill, EvoSkills, and EvoSkill study evolvable memories, trajectory-derived procedures, verifier-style repair, and automated skill discovery \citep{zhang2026memrl,zhang2026memskill,xia2026skillrl,ni2026trace2skill,zhang2026coevoskills,alzubi2026evoskill}. Evaluation work cautions that skill injection can be neutral or negative under context mismatch, gains degrade in large skill pools, and executor selection becomes a bottleneck in overlapping registries \citep{han2026sweskills,liu2026skillusagewild,zheng2026skillrouter}. SkillMAS inherits utility-aware retrieval and verifier-style skill repair, but its narrower contribution is using one retained verified-trace evidence set to constrain both skill evolution updates and MAS-organization edits. Direct protocol-matched comparisons to these frameworks are not yet available in the current harnesses, so we position them as conceptual neighbors rather than head-to-head baselines.

\paragraph{Agent infrastructure, governance, and deployment value.}
Recent systems work broadens agent research from isolated task success to the infrastructure and deployment economics around persistent agents. Externalization frames memory, skills, protocols, and harness engineering as coupled runtime infrastructure rather than independent add-ons \citep{zhou2026externalization}. Holos pushes this view toward web-scale agentic ecosystems with persistent agents, market-style orchestration, and value-cycle design \citep{nie2026holos}. SkillProbe studies the complementary governance problem: emerging skill marketplaces can create semantic-behavioral and combinatorial security risks that require scalable auditing \citep{guo2026skillprobe}. Agentic ROI further argues that agent systems should be evaluated not only by raw success but by value relative to time, interaction, and cost \citep{liu2025agenticroi}. SkillMAS fits this infrastructure-oriented line by making skill growth, executor organization, and verification share the same evidence surface, so adaptation is treated as a runtime systems problem rather than only a benchmark-score improvement.

\section{Conclusion}
\label{sec:conclusion}

We presented SkillMAS, a nonparametric framework for post-deployment specialization in LLM-based multi-agent systems. The central premise is that skill evolution and MAS organization should not be treated as independent adaptation targets: evolving reusable procedures changes what the system must route, maintain, and verify, while the MAS organization determines whether those procedures can be applied without excessive context pressure or responsibility ambiguity. SkillMAS addresses this coupling through a shared verified-trace evidence surface. Utility Learning assigns credit only to execution-supported skills and participating executors; bounded skill evolution consolidates reusable procedures without unfiltered library growth; and evidence-gated MAS restructuring changes executor boundaries only when retained failures and Executor Utility indicate a structural mismatch.

Across embodied manipulation, command-line OS workflows, and retail-service interaction, SkillMAS is competitive under the reported harnesses and provides process evidence for coupled adaptation. The ALFWorld transfer stress test is consistent with the paper's mechanism-level claim that skill libraries and MAS organization can become mismatched when evolved separately. The adaptation trajectories further show that specialization is not uniformly beneficial: in \(\tau\)-Bench Retail, the selected checkpoint improves through skill-utility adaptation while preserving single-agent system, suggesting that SkillMAS is still useful because it can choose not to expand the MAS.

\textbf{Limitation and Future Work.} As a first step toward coupled post-deployment adaptation, the current paper has several important limitations. First, the evidence is benchmark-local and protocol-dependent: the reported gains are tied to particular harnesses, prompts, model APIs, and checkpoint-selection procedures, so they should not yet be read as a domain-agnostic estimate of SkillMAS. Second, although the ALFWorld transplant stress test is consistent with the paper's mechanism-level claim, we do not yet isolate the causal contribution of each adaptation component with uniform multi-seed reruns, controlled frozen-target ablations, or formal significance tests. Third, the current archive does not yet provide complete token, latency, and cost accounting, so the practical efficiency and longer-run stability of coupled adaptation remain undercharacterized. Finally, the paper still abstracts away many low-level trajectory records in favor of representative process summaries, so longer-horizon stability and failure-mode auditing remain only partially characterized in the current presentation. Future work will therefore prioritize protocol-matched ablations, longer-horizon stability and cost analyses, and richer trace-level auditing that makes the evolution of skills and executor boundaries inspectable end to end.

\clearpage
\bibliographystyle{plainnat}
\bibliography{refs}

\clearpage
\appendix

\section{Supplementary Discussion}
\label{app:supplementary}

\subsection{Algorithmic Instantiation}
\label{app:algorithmic_instantiation}

Algorithm~\ref{alg:skillmas_round} records the concrete round-level contract used in this draft. The operators are deliberately non-parametric: the shared state is updated from verified traces, while benchmark-local heuristics enter only through retained-evidence construction, diagnosability, and restructuring predicates.

\begin{algorithm}[H]
\caption{SkillMAS round sketch}
\label{alg:skillmas_round}
\small
\begin{algorithmic}[1]
\Require Round state $X_r=(\mathcal{L}_r,\mathcal{A}_r,Q^s_r,Q^a_r,\mathcal{P}_r,\mathcal{K}_r)$
\Ensure Updated state $X_{r+1}$
\State Execute a batch with fixed $X_r$ and record verified traces $\mathcal{T}_r$
\State Update $Q^{s,+}_r$ only for invoked or execution-supported skills, and update $Q^{a,+}_r$ only for executors with verified traces
\State Retain repeated failures, near misses, reusable successes, and retrieval/execution mismatches as $\widetilde{\mathcal{T}}_r$
\State Convert each retained trace into at most one skill proposal
\State Consolidate proposals with \textsc{SkillEvolve} to create, refine, prune, hold in pool, or no-op at most once per implicated skill cluster
\State Build $H_r$ artifacts from retained failures, Executor Utility, skill overlap, and pending skill updates
\State Apply $G_r$ to keep, add, merge/remove, or modify one executor boundary only when the artifacts support a structural mismatch
\State Promote validated skills from $\mathcal{P}_r$, transfer owned skills if MAS organization changed, and form $X_{r+1}$
\end{algorithmic}
\end{algorithm}

\begin{table}[H]
\centering
\scriptsize
\caption{Benchmark-local heuristic instantiations used in the current experiments. These rules are implementation details of this study, not universal thresholds.}
\label{tab:heuristic_instantiations}
\begin{tabularx}{\textwidth}{@{}p{0.14\textwidth}p{0.20\textwidth}p{0.19\textwidth}p{0.19\textwidth}Y@{}}
\toprule
\textbf{Benchmark} & \textbf{Phase vocabulary} & \textbf{Used-skill support} & \textbf{Diagnosable repair set} & \textbf{$H_r$ / $G_r$ evidence} \\
\midrule
ALFWorld & search, manipulation, appliance use, global & Skill id invocation or verifier-matched guard/step pattern in the action trace & add guard, reorder step, tighten retrieval, split skill, hand off to structure & Task-family failure mass, low executor utility after skill repair, search/manipulation/appliance skill overlap, one specialist add or boundary hold per round \\
Lifelong OS Task & filesystem, policy grounding, text-log checking, global & Skill id invocation or shell-command/output pattern matching the skill's verifier checks & add guard, reorder command sequence, tighten retrieval, split skill, hand off to structure & Concentrated failures by OS operation family, utility gap between generalist and candidate specialist, owned-skill overlap, one policy/text-log specialist or boundary modification per round \\
$\tau$-Bench Retail & global retail case handling & Skill id invocation or tool-call sequence matching the skill's case-graph and payload checks & add guard, tighten retrieval, hold in pool, hand off to structure & Transaction-state coupling, manager utility, helper handoff failure mass, boundary-overlap risk; selected checkpoint keeps single-agent execution \\
\bottomrule
\end{tabularx}
\end{table}

\subsection{Experiment Configuration Summary}
\label{app:experiment_config}

Table~\ref{tab:experiment_config} summarizes the experimental settings needed to interpret the reported results. We keep this table at the benchmark-protocol level and include only reporting details that affect how the main-table scores should be read.

\begin{table}[H]
\centering
\small
\caption{Benchmark-level configuration summary.}
\label{tab:experiment_config}
\begin{tabularx}{\textwidth}{@{}p{0.18\textwidth}p{0.27\textwidth}p{0.22\textwidth}Y@{}}
\toprule
\textbf{Benchmark} & \textbf{Adaptation setting} & \textbf{Evaluation setting} & \textbf{Model roles} \\
\midrule
ALFWorld unseen domain \citep{shridhar2021alfworld} & Fixed 70-task train subset used for round-by-round adaptation tracking; see Appendix Section~\ref{app:subset_construction}. & Full unseen 134-task split; the best adaptation checkpoint is evaluated once for the headline result. & GPT-4o-mini is used wherever possible. \\
Lifelong Agent Bench OS Task domain \citep{zheng2025lifelongagentbench} & Fixed 100-task train subset used for process tracking; see Appendix Section~\ref{app:subset_construction}. & Full 150-task evaluation set; the headline result reports the best checkpoint. & GPT-4o-mini is used wherever possible. \\
$\tau$-Bench Retail domain \citep{yao2024taubench} & Official 74-task train subset used for adaptation and development. & The main table reports the benchmark base set with 114 tasks total: 74 train tasks plus 40 test tasks. & Fixed benchmark roles: GPT-4.1-mini executor and GPT-4.1-2025-04-14 user simulator. \\
\bottomrule
\end{tabularx}
\end{table}

For provenance, the unmarked Lifelong Agent Bench OS Task and $\tau$-Bench Retail baselines in Table~\ref{tab:main_results} are our evaluations or reruns under the current harness, whereas the ALFWorld unseen-domain entries marked ``Ref.'' remain published contextual references from source-paper protocols.

\subsection{Benchmark Train Set Construction}
\label{app:subset_construction}

We record the concrete subset construction rules here.

\textbf{ALFWorld unseen domain.} The adaptation trajectory is tracked on a fixed 70-task train subset selected once from the 134-task set by task-family-proportional sampling. The same fixed subset is reused across rounds so that changes in success, skill count, and executor count reflect adaptation rather than resampling noise.

\textbf{Lifelong Agent Bench OS Task domain.} The process table uses a fixed 100-task train subset selected once from the Lifelong Agent Bench OS Task for round-by-round tracking. The full 150-task evaluation score is reported separately from the best checkpoint, so the train subset is used only for process measurement rather than as the headline evaluation set.

\textbf{$\tau$-Bench Retail domain.} The official 74-task train subset is used for adaptation and development under the benchmark's fixed-role setup, while the main benchmark evaluation is the 114-task base set consisting of 74 train tasks and 40 test tasks. The benchmark fixes two model roles, a GPT-4.1-mini executor and a GPT-4.1-2025-04-14 user simulator, and the reported main-table values are read under that benchmark-defined interaction protocol rather than under the ALFWorld or Lifelong Agent Bench OS Task harness semantics.

\subsection{Round-by-Round Adaptation Trajectories}
\label{app:adaptation_trajectories}

Table~\ref{tab:process_rounds} reports the process summaries used to interpret how the adapted state evolves before the headline evaluation checkpoints.

\begin{table}[H]
\centering
\small
\setlength{\tabcolsep}{4pt}
\caption{Round-by-round adaptation summaries. Skills and Executors denote active counts.}
\label{tab:process_rounds}
\begin{tabular}{@{}c r c c p{0.30\textwidth}@{}}
\toprule
\multicolumn{5}{l}{\textbf{ALFWorld unseen domain (70-task train subset)}} \\
\midrule
\textbf{R} & \textbf{Success} & \textbf{Skills} & \textbf{Executors} & \textbf{Executor state} \\
\midrule
0 & 54/70 (77.1\%) & 5 & 2 & seed executors \\
1 & 61/70 (87.1\%) & 7 & 3 & + search executor \\
2 & 62/70 (88.6\%) & 9 & 4 & + manipulation executor \\
3 & 63/70 (90.0\%) & 11 & 4 & + appliance executor \\
4 & 62/70 (88.6\%) & 11 & 4 & stable executors \\
5 & 66/70 (94.3\%) & 13 & 4 & stable executors \\
\midrule
\multicolumn{5}{l}{\textbf{Lifelong Agent Bench OS Task domain (100-task train subset)}} \\
\midrule
\textbf{R} & \textbf{Success} & \textbf{Skills} & \textbf{Executors} & \textbf{Executor state} \\
\midrule
0 & 62/100 (62.0\%) & 4 & 2 & seed executors \\
1 & 65/100 (65.0\%) & 7 & 2 & stable executors \\
2 & 75/100 (75.0\%) & 9 & 3 & + policy-grounding executor \\
3 & 78/100 (78.0\%) & 10 & 4 & + text-log executor \\
4 & 75/100 (75.0\%) & 10 & 4 & stable executors \\
\midrule
\multicolumn{5}{l}{\textbf{$\tau$-Bench Retail domain (74-task train subset; selected trajectory plus expansion probe)}} \\
\midrule
\textbf{R} & \textbf{Success} & \textbf{Skills} & \textbf{Executors} & \textbf{Executor state} \\
\midrule
0 & 43/74 (58.1\%) & 6 & 1 & seed executor \\
1 & 51/74 (68.9\%) & 6 & 1 & stable executor \\
2 & 32/74 (43.2\%) & 10 & 2 & + preflight executor \\
\bottomrule
\end{tabular}
\end{table}

\subsection{ALFWorld Executor Prompt and Skill Snapshot}
\label{app:alfworld_case}

For interpretability, we summarize ALFWorld specialization as a coupled adaptation reconstruction of the evolved SkillMAS lineage. An executor and its skill library are treated as one unit: adding a worker is meaningful only when the executor receives reusable procedures that make the role operational.

\subsubsection{Executor Prompt Cards}

\begin{agentcard}{General Executor}
\small

\textbf{Prompt excerpt.} The manager acts as an orchestration layer: parse the household task, delegate the smallest safe next subtask, and carry forward exact object and receptacle state instead of guessing.

\textbf{Skill library.} \texttt{alfworld/task\_decomposition}, \texttt{alfworld/domain\_knowledge}, and \texttt{alfworld/pick\_and\_place\_procedure}.

\textbf{Low-value neighbors.} Early RL-style patches for examine and cleaning tasks were counted in the raw bank but did not become stable reusable procedures.

\textbf{Coordination logic.} The general executor establishes the initial manager/generalist contract before any specialist split is useful; skills are broad and executor assignment remains simple.
\end{agentcard}

\medskip

\begin{agentcard}{Search Executor}
\small

\textbf{Prompt excerpt.} The search worker is responsible for finding the exact object or destination, reading observations after each \texttt{goto}, opening closed containers when needed, and stopping once exact evidence is found.

\textbf{Skill library.} \texttt{alfworld/object\_search\_strategy}, \texttt{alfworld/examine\_procedure}, and \texttt{pick\_cool\_then\_place\_learned\_r0}.

\textbf{Low-value neighbors.} Prompt-like examine summaries and generic placement RL patches were rejected because they repeated local trajectory wording without transferable search guards.

\textbf{Coordination logic.} Search is separated from execution so later workers receive grounded object and receptacle names rather than open-ended navigation instructions.
\end{agentcard}

\medskip

\begin{agentcard}{Manipulation Executor}
\small

\textbf{Prompt excerpt.} The manipulation worker executes exact take/put operations, preserves numbered object names, opens only confirmed closed containers, and diagnoses placement failures through location, openness, and holding state.

\textbf{Skill library.} \texttt{alfworld/object\_handling}, \texttt{alfworld/pick\_and\_place\_procedure}, and \texttt{pick\_cool\_then\_place\_learned\_r2}.

\textbf{Low-value neighbors.} Generic pick-and-place summaries and two-object RL placeholders were rejected because they did not add exact-reference discipline beyond the skill library.

\textbf{Coordination logic.} Once search has grounded the object, manipulation can focus on physical state transitions rather than rediscovering locations or inventing alternate destinations.
\end{agentcard}

\medskip

\begin{agentcard}{Appliance Executor}
\small

\textbf{Prompt excerpt.} The appliance worker handles heat, cool, clean, and lamp-toggle steps after the object and appliance are grounded; it navigates to the exact appliance, performs one operation, and reports concrete evidence.

\textbf{Skill library.} \texttt{alfworld/appliance\_operation}, \texttt{alfworld/pick\_and\_place\_optimization}, and \texttt{alfworld/examine\_optimization}.

\textbf{Low-value neighbors.} Late examine and heat summaries that stayed prompt-like were rejected; heating and two-object skills were accepted only when they complemented the appliance executor with concrete procedure.

\textbf{Coordination logic.} The final executor removes transformation logic from the generalist path, so heating, cooling, cleaning, and lamp-based examine tasks are handled as appliance operations rather than ad hoc placement variants.
\end{agentcard}

\subsubsection{Representative Skill Cards}

The appendix does not reproduce every skill file inline. It shows representative validated skills that anchor the full-tree narrative and summarizes nearby low-value artifacts as rejection evidence.

\begin{skillcard}{\texttt{alfworld/task\_decomposition} (manager seed skill)}
\small
\textbf{When to use.} Whenever the manager receives a new ALFWorld task and must decide the next worker-safe substep.

\textbf{Accepted reasoning.} Decompose by task family, then carry exact object, source, destination, holding state, and openness across delegated steps.

\textbf{Representative steps.}
\begin{enumerate}[leftmargin=1.25em,itemsep=1pt,topsep=2pt]
\item Identify task family: place, clean, heat, cool, or examine.
\item Route search before manipulation when the exact object or receptacle is unknown.
\item Route appliance use only after the exact object and exact appliance are grounded.
\end{enumerate}

\textbf{Why kept / why rejected.} Kept because a worker split without manager-side decomposition would not form a coherent MAS story. Rejected neighbors are early placeholder files that still treat the task as one undifferentiated workflow.
\end{skillcard}

\medskip

\begin{skillcard}{\texttt{alfworld/object\_search\_strategy} (search skill)}
\small
\textbf{When to use.} Any search step that must resolve an exact object or exact destination receptacle.

\textbf{Accepted reasoning.} Search open surfaces first, then verified closed containers, and stop as soon as exact evidence is available.

\textbf{Representative steps.}
\begin{enumerate}[leftmargin=1.25em,itemsep=1pt,topsep=2pt]
\item Classify the target category to choose high-probability locations.
\item Use exact environment names such as \texttt{cabinet 5} or \texttt{countertop 1}.
\item Open only the confirmed closed receptacle that is currently under inspection.
\end{enumerate}

\textbf{Why kept / why rejected.} Kept because it turns noisy search traces into a reusable executor rule. Rejected neighbors mixed object search, take actions, and lamp operations inside one prompt-like patch.
\end{skillcard}

\medskip

\begin{skillcard}{\texttt{alfworld/object\_handling} (manipulation skill)}
\small
\textbf{When to use.} Exact take or put steps after the manager has already grounded the target object and receptacle.

\textbf{Accepted reasoning.} Placement errors should be debugged through exact names, openness, and holding state, not through speculative alternate destinations.

\textbf{Representative steps.}
\begin{enumerate}[leftmargin=1.25em,itemsep=1pt,topsep=2pt]
\item Go to the exact source or destination.
\item Open the exact container only if it is confirmed closed.
\item Take or put using the exact object and receptacle names, then verify holding state if needed.
\end{enumerate}

\textbf{Why kept / why rejected.} Kept because the same manipulation invariants recur across task families. Rejected neighbors remained broad placement summaries without stronger exact-reference safeguards.
\end{skillcard}

\medskip

\begin{skillcard}{\texttt{alfworld/appliance\_operation} (appliance skill)}
\small
\textbf{When to use.} Heat, cool, clean, and lamp-toggle steps once the object and appliance are already grounded.

\textbf{Accepted reasoning.} Appliance use should be expressed as exact appliance navigation plus one explicit operation; the examine mechanic is the same rule specialized to lamp toggle while holding the object.

\textbf{Representative steps.}
\begin{enumerate}[leftmargin=1.25em,itemsep=1pt,topsep=2pt]
\item Go to the exact appliance or lamp.
\item Open it only if the operation requires that state.
\item Perform the requested operation and report concrete evidence of success or failure.
\end{enumerate}

\textbf{Why kept / why rejected.} Kept because it absorbs brittle transformation rules into one reusable executor role. Rejected neighbors never became appliance-specific enough to retain.
\end{skillcard}

\medskip

\begin{skillcard}{\texttt{pick\_heat\_then\_place\_learned\_r5} (accepted late trajectory skill)}
\small
\textbf{When to use.} Stabilized heat-transform-place tasks after the appliance executor is already in place.

\textbf{Accepted reasoning.} Search target object, retrieve it, move to the exact microwave or stove, heat it, then place it at the grounded destination.

\textbf{Representative steps.}
\begin{enumerate}[leftmargin=1.25em,itemsep=1pt,topsep=2pt]
\item Locate and pick up the target item.
\item Navigate to the exact heating appliance and perform the heat operation.
\item Carry the heated item to the exact destination.
\end{enumerate}

\textbf{Why kept / why rejected.} Kept because it complements the appliance executor with a concrete multi-step heating routine. Earlier heating summaries were rejected when they stayed generic or lacked verification.
\end{skillcard}

\medskip

\begin{skillcard}{\texttt{pick\_two\_obj\_and\_place\_learned\_r5} (accepted late trajectory skill)}
\small
\textbf{When to use.} Multi-instance placement tasks where the system must distinguish exact object references and keep track of what has already been collected.

\textbf{Accepted reasoning.} Multi-object tasks require exact instance references, container-state management, and explicit tracking of which object has already been placed.

\textbf{Representative steps.}
\begin{enumerate}[leftmargin=1.25em,itemsep=1pt,topsep=2pt]
\item Locate both target instances with exact names.
\item Open containers before retrieval and keep exact references during transport.
\item Place both objects into the exact destination while verifying completion after each placement.
\end{enumerate}

\textbf{Why kept / why rejected.} Kept because it adds exact-reference discipline that the seed single-object routines did not encode. Earlier two-object patches were rejected when they lacked collection-state tracking.
\end{skillcard}

\subsection{Lifelong Agent Bench OS Task Executor Prompt and Skill Snapshot}
\label{app:os-snapshot}

The OS-task snapshot comes from the round-3 best checkpoint of the Lifelong Agent Bench OS Task domain run. The best train-subset score is 78/100, with a transfer check of 63/100 on the next 100 training tasks. We include this snapshot to illustrate the executor boundaries and retained case skills behind the reported OS Task trajectory.

\subsubsection{Executor Prompt Cards}

\begin{agentcard}{General Executor}
\small
\textbf{Executor.} \texttt{general\_worker}, renamed in the prompt as \texttt{FilesystemSymlinkWorker}.

\textbf{Prompt excerpt.} The worker handles short Ubuntu command-line tasks end to end, must call \texttt{env.bash} first, must avoid \texttt{sudo} and interactive commands, and should finish only after lightweight checks with \texttt{test}, \texttt{stat}, \texttt{getent}, \texttt{grep}, \texttt{wc}, or \texttt{readlink}.

\textbf{Responsibility boundary.} Covers filesystem structure, symlinks, permissions, and file/group ownership changes. It may retain fallback text/log knowledge for mixed tasks, but does not specialize in text/log analysis after round-3 MAS restructuring.

\textbf{Representative skills.} Symlink target permissions, directory structure creation, group-owned file setup, and validator-script style checks.
\end{agentcard}

\medskip

\begin{agentcard}{Policy-Grounding Executor}
\small
\textbf{Executor.} \texttt{user\_group\_policy\_worker}, added by bounded MAS restructuring before the best checkpoint.

\textbf{Prompt excerpt.} The worker handles user, group, account-policy, membership, setgid, and group-owned workspace tasks. It uses exact paths, users, groups, modes, and output formats, and verifies state with commands such as \texttt{getent}, \texttt{id}, \texttt{stat}, and \texttt{chage -l}.

\textbf{Responsibility boundary.} Choose this worker when the task mentions group creation, user addition, group membership, setgid, directory group ownership, or explicit permission setting.

\textbf{Representative skills.} Group creation plus user membership, group-owned directory setup, and \texttt{gpasswd}-based administrator assignment.
\end{agentcard}

\medskip

\begin{agentcard}{Text-Log Executor}
\small
\textbf{Executor.} \texttt{text\_log\_count\_worker}, created in round 3 after text/log/counting skills accumulated inside the general worker.

\textbf{Prompt excerpt.} The worker handles text processing, log analysis, line counting, exact report filenames, and machine-checked output contents. It is routed to tasks involving \texttt{grep}, \texttt{awk}, \texttt{sed}, \texttt{wc}, real newlines, and exact-output files.

\textbf{Responsibility boundary.} Avoid filesystem structure, symlink, permission, or user/group management unless those operations directly support a text/log output requirement.

\textbf{Representative skills.} Generating exact log reports and summing lines across \texttt{.log} files without relying on misleading multi-file \texttt{wc -l} totals.
\end{agentcard}

\subsubsection{Representative Skill Cards}

The round-3 OS snapshot contains 10 active case-learned skills. We show three representative cards below, one for each major boundary in the evolved worker organization, rather than reproducing every retained skill file.

\begin{skillcard}{Group membership and group-owned workspace setup}
\small
\textbf{Source skill.} \texttt{case\_r0\_ep\_15\_create\_group\_add\_users...}

\textbf{When to use.} When a task asks the system to create a group, add users, set group ownership and permissions on directories or files, and write a human-readable membership log.

\textbf{Accepted reasoning.} Use \texttt{groupadd -f}, create missing users idempotently, add users with \texttt{gpasswd} or \texttt{usermod -aG}, then set group ownership and permissions on both directories and files before writing evaluator-visible logs.

\textbf{Critical don'ts.} Do not dump raw \texttt{getent group} output when the task asks for a readable report; do not forget to set both directory and file metadata.
\end{skillcard}

\medskip

\begin{skillcard}{Symlink target permissions and exact link paths}
\small
\textbf{Source skill.} \texttt{case\_r0\_ep\_23\_create\_directory\_structure...}

\textbf{When to use.} When a task combines directory creation, file creation, absolute or relative symlinks, and permissions on the linked target file.

\textbf{Accepted reasoning.} Create parent directories first, create the target file, construct absolute and relative symlinks with exact targets, and apply \texttt{chmod} to the target file rather than to the symlink itself.

\textbf{Verification.} Check \texttt{readlink -f} for absolute links, \texttt{readlink} for relative links, \texttt{stat -c '\%a'} for modes, and \texttt{cat} for requested file contents.
\end{skillcard}

\medskip

\begin{skillcard}{Log-line counting without temporary files}
\small
\textbf{Source skill.} \texttt{case\_r3\_ep\_29\_sum\_lines\_in\_all\_log\_files...}

\textbf{When to use.} When a task asks for the total number of lines across all \texttt{.log} files in a directory and requires writing the result to a single output file.

\textbf{Accepted reasoning.} Avoid \texttt{wc -l file1 file2 ...} when the total row can confuse downstream counting. Use \texttt{cat /var/log/app/*.log | wc -l} when matching files exist, and write \texttt{0} when no log files exist.

\textbf{Validation signal.} The repair script for the source failure passed static contract lint and source-episode benchmark validation on the first attempt.
\end{skillcard}

\subsection{\texorpdfstring{$\tau$-Bench}{Tau-Bench} Single-Agent Case Study}
\label{app:taubench-snapshot}

The $\tau$-Bench snapshot comes from the selected checkpoint-state evaluation summarized in Table~\ref{tab:process_rounds}. Telemetry shows no promoted scaling-helper usage in the selected trajectory, so the appendix records the manager-dominant prompt and skill state rather than a larger active MAS.

The post-selection expansion probe is documented in Table~\ref{tab:process_rounds}; the prompt card below records only the selected active system.

\begin{agentcard}{General Executor}
\small
\textbf{Executor set.} Single manager only.

\textbf{Prompt excerpt.} The manager is the single-agent case handler: authenticate, maintain a compact case graph, ground order status and item ids, choose the correct retail tool, execute confirmed writes directly, and close only after all asks are resolved.

\textbf{Skill library.} \texttt{founder\_case\_graph}, \texttt{identity\_order\_grounding}, \texttt{catalog\_variant\_selection}, \texttt{payload\_preflight}, \texttt{transaction\_execution}, and \texttt{closure\_audit}.

\textbf{Low-value neighbors.} Read-only grounding, catalog, and preflight helpers were available in the broader tree, but they were not part of the selected active manager path.

\textbf{Coordination logic.} $\tau$-Bench retail workflows bind identity, order status, item ids, payment history, write ordering, and final closure into one transactional state. The selected system therefore keeps these decisions with a single manager and uses Skill Utility to improve which seed procedures are applied, rather than splitting responsibility across active executors.
\end{agentcard}

\subsubsection{Representative Skill Cards}

The selected $\tau$-Bench snapshot contains six active seed skills. We show the cards that explain the manager-dominant result: exact case-graph maintenance, account/order grounding, variant selection, payload validation, mutation execution, and final closure.

\begin{skillcard}{\texttt{founder\_case\_graph} (manager seed skill)}
\small
\textbf{When to use.} Every retail interaction that requires the manager to combine account identity, order state, requested changes, confirmations, writes, and final closure.

\textbf{Accepted reasoning.} Keep one internal case graph with authenticated user, relevant orders, current item ids, candidate item ids, planned writes, completed writes, blockers, and open questions.

\textbf{Representative steps.}
\begin{enumerate}[leftmargin=1.25em,itemsep=1pt,topsep=2pt]
\item Ground the user and plausible orders before asking for more identifiers.
\item Solve directly when the next write is clear, confirmed, and status-compatible.
\item Treat helper output as evidence and reconcile exact ids before any mutation or final reply.
\end{enumerate}

\textbf{Why kept / why rejected.} Kept because $\tau$-Bench rewards continuous single-agent case handling; splitting responsibility too early increased routing and closure risk in later rounds.
\end{skillcard}

\medskip

\begin{skillcard}{\texttt{identity\_order\_grounding} (grounding skill)}
\small
\textbf{When to use.} Authentication, order scans, order-status gating, and current item mapping.

\textbf{Accepted reasoning.} Authenticate once, inspect user details, scan plausible orders, and label each requested target as pending-lane, delivered-lane, readonly-only, or blocked.

\textbf{Representative steps.}
\begin{enumerate}[leftmargin=1.25em,itemsep=1pt,topsep=2pt]
\item Prefer concrete credentials already present in the task.
\item Build an order map with order id, status, current item ids, product ids, address, and payment methods.
\item Separate pending, delivered, processed, canceled, and unknown-status orders before choosing a tool lane.
\end{enumerate}

\textbf{Why kept / why rejected.} Kept because most downstream failures start from wrong order or status grounding. Rejected alternatives asked for identifiers prematurely or restarted authentication after the user was already grounded.
\end{skillcard}

\medskip

\begin{skillcard}{\texttt{catalog\_variant\_selection} (catalog skill)}
\small
\textbf{When to use.} Product details, replacement variants, availability counts, exact candidate item ids, and price or refund inputs.

\textbf{Accepted reasoning.} Replacement ids must be real available variants of the current item's product; requested attributes are hard filters before price or similarity preferences.

\textbf{Representative steps.}
\begin{enumerate}[leftmargin=1.25em,itemsep=1pt,topsep=2pt]
\item Resolve the current product id from order details before variant lookup.
\item Filter variants by confirmed attributes and \texttt{available=true}.
\item Preserve unchanged attributes for ``same but'' or ``similar'' requests and return one exact candidate only when evidence supports it.
\end{enumerate}

\textbf{Why kept / why rejected.} Kept because wrong item-id and near-miss replacement choices dominated hard exchange-return failures. Rejected behavior used product ids as item ids or selected unavailable near matches.
\end{skillcard}

\medskip

\begin{skillcard}{\texttt{payload\_preflight} (write-guard skill)}
\small
\textbf{When to use.} Before return, exchange, cancel, pending modify, payment update, address update, or any broad multi-item mutation.

\textbf{Accepted reasoning.} A safe payload must match the order status, use current item ids from the target order, use available same-product candidate item ids where required, ground payment methods, and cover the full confirmed scope.

\textbf{Representative steps.}
\begin{enumerate}[leftmargin=1.25em,itemsep=1pt,topsep=2pt]
\item Check pending-only and delivered-only tool lanes before execution.
\item Verify array alignment and item-id provenance for current and replacement items.
\item Block same-order incompatible mutation mixes unless the user has given an explicit priority choice.
\end{enumerate}

\textbf{Why kept / why rejected.} Kept because it converts many benchmark-specific mistakes into general status, id, payment, and confirmation checks. Rejected variants became prompt patches for individual stories rather than reusable mutation guards.
\end{skillcard}

\medskip

\begin{skillcard}{\texttt{transaction\_execution} (manager-run mutation skill)}
\small
\textbf{When to use.} One grounded $\tau$-Bench write after the manager has exact tool arguments and confirmation or policy permission.

\textbf{Accepted reasoning.} Execute one write, then update the case graph with affected order id, changed item ids, address or payment effects, refund or price facts, and remaining uncertainty.

\textbf{Representative steps.}
\begin{enumerate}[leftmargin=1.25em,itemsep=1pt,topsep=2pt]
\item Verify the planned write names exactly one tool and exact arguments.
\item Re-check status or item provenance when the case graph is incomplete.
\item Execute the write and use the tool result, not an old plan, for the customer summary.
\end{enumerate}

\textbf{Why kept / why rejected.} Kept as a manager-run procedure because delegated transaction execution did not produce stable promoted-helper evidence in the selected trajectory.
\end{skillcard}

\medskip

\begin{skillcard}{\texttt{closure\_audit} (closure skill)}
\small
\textbf{When to use.} Before final reply, transfer, or \texttt{done}.

\textbf{Accepted reasoning.} Close only when all factual asks are answered, confirmed writes are complete or concretely blocked, broad scopes are explicitly covered, and final text names concrete order, item, payment, refund, or price-difference facts.

\textbf{Representative steps.}
\begin{enumerate}[leftmargin=1.25em,itemsep=1pt,topsep=2pt]
\item Verify each confirmed write has a matching completed write or blocker.
\item Check multi-order and multi-item scopes for skipped rows.
\item Use latest tool results for status, payment, refund, balance, and price-difference statements.
\end{enumerate}

\textbf{Why kept / why rejected.} Kept because many failures were completed writes with incomplete or looping closure. Rejected neighbors kept reopening already-completed work or promised unsupported follow-up.
\end{skillcard}

\end{document}